\documentclass[a4paper,12pt]{article}
\usepackage{times}
\usepackage{epsfig}
\usepackage{epstopdf}
\usepackage{epsfig}
\usepackage{ccaption}
\usepackage{t1enc,a4}
\usepackage{amssymb}
\usepackage{cite}
\usepackage[sumlimits,intlimits,namelimits]{amsmath} 
\usepackage[english]{babel}
\usepackage{graphicx}
\usepackage{geometry}

\numberwithin{equation}{section}
\newcommand{\bra}[1]{\left|#1\,\right\rangle}

\newcommand{\mytab}[4][abcd\thetables]{\refstepcounter{tables}%
                       \begin{table}[hbt]%
		       \begin{center}%
		       \begin{tabular}{#2}%
		       #3%
		       \end{tabular}%
		       \end{center}%
		       \caption{\label{#1}\small #4}
		       \end{table}}

\captionnamefont{\small\bfseries}
\captiontitlefont{\small\mdseries}
\captiondelim{:\,\,\,}
\hangcaption

\newcounter{figures}
\newcounter{tables}
\newcommand{\us}{$U$-spin }
\newcommand{\usd}{$U$-spin}

\begin{document}
\begin{titlepage}
\begin{flushright}
SI-HEP-2009-01 \\[0.2cm]
\today
\end{flushright}

\vspace{1.2cm}
\begin{center}
{\Large\bf 
General Analysis of \boldmath $U$-Spin Breaking in $B$ Decays\unboldmath}
\end{center}

\vspace{0.5cm}
\begin{center}
{\sc Martin Jung$^{a,b}$ and Thomas Mannel$^a$} \\[0.1cm]
{\sf $^a$ Theoretische Physik 1, Fachbereich Physik,
Universit\"at Siegen\\ D-57068 Siegen, Germany\\[2ex]
$^b$ Instituto de F\'{\i}sica Corpuscular (IFIC),
CSIC-Universitat de Val\`encia, \\
Apartado de Correos 22085,
E-46071 Valencia, Spain.
}
\end{center}

\vspace{0.8cm}
\begin{abstract}
\vspace{0.2cm}\noindent
We analyse the breaking of \us on a group theoretical basis. Due to the simple 
behaviour of the weak effective hamiltonian under \us and the unique structure 
of the breaking terms such a group theoretical analysis leads to a manageable number 
of parameters. Several applications are discussed, including the decays $B \to J/\psi K$ 
and $B \to D K$.
\end{abstract}

\end{titlepage}

\newpage
\pagenumbering{arabic}

\section{Introduction}
Non-leptonic decays of bottom hadrons play an important role in the investigation  
of CP violation. While this effect has been established in non-leptonic Kaon decays, and
the CKM mechanism of CP violation is consistent with what is seen in non-leptonic $B$ 
meson decays, the observations cannot be easily linked with the fundamental parameters.  

The reason for this is the lack of a reliable method to compute the amplitudes of 
non-leptonic decays, which are given by matrix elements of an effective interaction 
expressed in terms  of a combination of four-quark operators between meson states. 
The QCD dynamics turn out to be so complicated, that currently neither 
factorization based methods nor lattice calculations yield reliable 
and precise predictions. 

While non-leptonic decays are a nice laboratory for studying QCD methods, the road to 
precise predictions for CP violation in non-leptonic $B$ decays seems to be the use of flavour 
symmetries, supplemented by the enormous amount of data expected from LHC and the (Super)
flavour factories. From the current perspective this will remain true for some time, until 
a qualitative breakthrough is achieved in the field of QCD methods.

There is a vast literature in the field of flavour symmetries and their applications to $B$ decays. 
The complete decomposition of the two-body $B$ decay amplitudes in terms of irreducible 
flavour $SU(3)$ matrix elements has been performed in \cite{GrinsteinLebed,Zeppenfeld}. 
Applications to $B$ decays have been considered in 
\cite{Gronau:1994rj,Dighe:1995gq,GronauUSpin,Gronau:2000ru,Gronau:2003ep,SavageWise}, and we 
shall use the notation of \cite{SoniSuprun}. Furthermore, 
flavour-symmetry strategies related to the extraction of CP violating parameters have been 
discussed in 
\cite{SoniSuprun,Gronau:1994bn,Buras:1994pb,Buras:1998rb}.  
 
The main problem with flavour symmetries is that they hold only approximately.  The usual starting 
point is flavour $SU(3)$, which suffers from a substantial breaking due to the sizable mass of the 
strange quark \cite{Gronau:1995hm}. On the other hand, the isospin subgroup is 
known to have a much smaller breaking, and 
can be safely assumed to be unbroken in this context. Hence one may consider the other two possibilities to 
identify $SU(2)$ subgroups of flavour $SU(3)$, which run under the names \us and 
$V$-spin.
Among these two subgroups,  the generators of \usd, under which the $d$ and the $s$ 
quark form a fundamental doublet, commute with the charge operator, which makes this 
subgroup particularly interesting with respect to electroweak interactions.  On the other 
hand, \us symmetry is broken at the same level as the full flavour $SU(3)$ due to the
splitting $m_s - m_d$. 

However, this breaking has a simple structure and can be readily included by a spurion 
analysis. In the present paper we perform such an analysis and apply it to non-leptonic
$B$ decays. We make use of present data to extract the symmetry-breaking matrix elements 
and give relations which include \us breaking, which can be tested in the future, 
once more data are available.   

\section{Flavour Symmetries}
\us is an $SU(2)$ subgroup of the full  $SU(3)$ flavour symmetry group, in which 
the $d$ and the $s$ quark form a doublet. A priori, \us is as badly broken as the full $SU(3)$, 
since the masses of the two quarks are substantially different:
$$
\Delta m \equiv m_s - m_d \sim \Lambda_{\rm QCD}\,,
$$
where $\Lambda_{\rm QCD} $ denotes the nonperturbative QCD scale. Regarding the group structure
of the breaking term, we observe that the relevant  mass term in the Lagrangian reads
\begin{eqnarray}
{\cal L}_{s\,\rm mass} &=& m_d \bar{d} d + m_s  \bar{s} s  \nonumber 
  = \frac{1}{2}  (m_s + m_d) (\bar{d} d  +  \bar{s} s) +  \frac{1}{2}  \Delta m  (\bar{s} s  -  \bar{d} d)  \\
                                 &=& \frac{1}{2}  (m_s + m_d) \bar{q} q  -  \frac{1}{2}  \Delta m  \bar{q} \tau_3 q  \,,
\end{eqnarray}  
where 
$$
q = \left( \begin{array}{c} d \\ s \end{array} \right) 
$$
is the \us quark doublet. Thus we conclude that the breaking term can be described as a triplet 
spurion 
\begin{equation} 
{\cal H}_{\rm break} =  \frac{1}{2}  \Delta m   \tau_3   = \epsilon B^{(1)}_0 \,,
\end{equation} 
where $B^{(1)}_0$ is an irreducible tensor-operator with $j = 1$ and 
$j_3 = 0$ of \usd. Here we also introduce the  small quantity $\epsilon$ related to the symmetry 
breaking   $\epsilon  \sim  \Delta m / \Lambda_{\chi SB}$ where  $ \Lambda_{\chi SB}$ 
is the chiral symmetry breaking scale.  

If we consider a matrix element of some operator ${\cal O} (x)$, which can be decomposed into 
irreducible tensor-operators of \usd, we may consider \us breaking to leading order 
by evaluating 
\begin{equation} \label{break} 
 \langle \tilde{f}  | {\cal O} (0)  | \tilde{i} \rangle  
=  \langle  f  | {\cal O} (0) |  i \rangle  + (-i) \int d^4 x \, 
          \langle  f  | T [ {\cal O} (0) \mathcal{H}_{\rm break} (x)]  |  i \rangle + \ldots  \,,
\end{equation} 
where the states $\tilde{f}$ and $\tilde{i}$ include the breaking term, while the states 
$f$ and $i$ are the \us symmetric states.   A general analysis of \us breaking 
can be perfomed by a group theory analysis of the breaking term by  decomposing the 
$T$ product of the operator ${\cal O}$ with $\mathcal{H}_{\rm break}$ into irreducible 
tensor operators $T^{(j)}_{j_3}$ of \usd. 

The simplest non-trivial case emerges if the operator ${\cal O}$ is an \us  doublet, 
which we denote by ${\cal O}^{(1/2)}_{j_3}$.  In this case, the last term in  (\ref{break}) 
decomposes into   
\begin{eqnarray}  
(-i) \int d^4 x \,  T\!\left[ {\cal O}^{(1/2)}_{\pm1/2} (0) \mathcal{H}_{\rm break} (x) \right]  
&=&   (-i \epsilon ) \int d^4 x \,  T\!\left[ {\cal O}^{1/2}_{\pm1/2} (0) B^{(1)}_0 (x) \right]   \nonumber \\ 
&=&   \sqrt{\frac{2}{3}} \left[ K^{(3/2)}_{\pm1/2} \mp \sqrt{\frac{1}{3}} K^{(1/2)}_{\pm1/2} \right]\,.
\end{eqnarray} 

Aside from the trivial example  of  the currents $j = \bar{u} \Gamma q$, $q= d,s$ ,
also  the effective weak hamiltonian  for $B$ decays is a pure \us doublet, 
even if electroweak penguins are included. The latter is true due to the fact that 
the $s$ and the $d$ quark carry the same electroweak quantum numbers. 
Thus from the group theoretical point of view we may decompose the weak 
effective hamiltonian into its irreducible tensor components according to 
\begin{eqnarray}\label{IrrOp}
H_{\rm eff}^{\Delta C = \pm 1} &=& \frac{4 G_F}{\sqrt{2}} \left[ V_{cb} V_{ud}^* P^{(1/2)}_{1/2} 
+  V_{cb} V_{us}^* P^{(1/2)}_{-1/2} \right]\,, \\
H_{\rm eff}^{\Delta C = 0} &=& \frac{4 G_F}{\sqrt{2}} 
\left[ V_{cb} V_{cd}^* Q^{(1/2)}_{1/2} + V_{ub} V_{ud}^* R^{(1/2)}_{1/2} + 
V_{cb} V_{cs}^* Q^{(1/2)}_{-1/2} + V_{ub} V_{us}^* R^{(1/2)}_{-1/2} \right]\,,
\end{eqnarray} 
where the operators $P^{(1/2)}_{j_3}$, $Q^{(1/2)}_{j_3}$, and $R^{(1/2)}_{j_3}$ are 
renormalization group invariant combinations of four-quark operators. 

In the following we shall use this group theoretical decomposition to discuss \us and 
its breaking in various $B$ decays.  
To this end, we have to identify the \us multiplets of hadronic states. Starting from the 
definition of the fundamental quark doublets 
(we use the same sign convention as in \cite{SoniSuprun}),
\begin{equation}
\left[\begin{array}{l}
|d\rangle \\
|s\rangle \end{array}\right]
=
\left[\begin{array}{l}
|\frac{1}{2}\,+\frac{1}{2}\rangle\\
|\frac{1}{2}\,-\frac{1}{2}\rangle\\
\end{array}\right]\,,\quad
\left[\begin{array}{l}
|\bar{s}\rangle \\
|\bar{d}\rangle \end{array}\right]
=
\left[\!\begin{array}{l}
\phantom{-}|\frac{1}{2}\,+\frac{1}{2}\rangle\\
-|\frac{1}{2}\,-\frac{1}{2}\rangle\\
\end{array}\right]\,,
\end{equation}
we obtain for the decaying $B$ mesons 
\begin{equation}
\bra{B^+} = \bra{(u \bar{b})}= \bra{0,0}  \, , \quad 
\left[\begin{array}{l}
\bra{B^0}=\bra{(d\bar{b})}\\
\bra{B_s}=\bra{(s\bar{b})}
\end{array}\right] = 
\left[\begin{array}{l}
\bra{\frac{1}{2},+\frac{1}{2}}\\ \bra{\frac{1}{2},-\frac{1}{2}}
\end{array}\right] \, . 
\end{equation}
The mesons in the final state are in terms of \us
\begin{eqnarray} \label{conv}
\left[\begin{array}{l}
\bra{K^+}=\bra{(u\bar{s})}\\
\bra{\pi^+}=\bra{(u\bar{d})}
\end{array}\right]&=&
\left[\begin{array}{l}
\phantom{-}\bra{\frac{1}{2},+\frac{1}{2}}\\-\bra{\frac{1}{2},-\frac{1}{2}}
\end{array}\right]\,,\\
\left[\begin{array}{l}
\bra{\pi^-}=-\bra{(\bar{u}d)}\\
\bra{K^-}=-\bra{(\bar{u}s)}
\end{array}\right]&=&
\left[\begin{array}{l}
-\bra{\frac{1}{2},+\frac{1}{2}}\\-\bra{\frac{1}{2},-\frac{1}{2}}
\end{array}\right]\,,\\
\left[\begin{array}{l}
\bra{K^0}=\bra{(\bar{s}d)}\\
\sqrt{3}/2\bra{\eta_8}-1/2\bra{\pi^0}=\bra{(\bar{s}s-\bar{d}d)}\\
\bra{\bar{K}^0}=\bra{(\bar{d}s)}
\end{array}\right]&=&
\left[\begin{array}{l}
\phantom{-}\bra{1,+1}\\
\phantom{-}\bra{1,\phantom{-}0}\\
-\bra{1,-1}
\end{array}\right]\,,\\
\left[\begin{array}{l}
\bra{K^{*0}}=\bra{(\bar{s}d)}\\
1/\sqrt{2}\bra{\phi}-1/2\bra{\rho^0}-1/2\bra{\omega}=\bra{(\bar{s}s-\bar{d}d)}\\
\bra{\bar{K}^{*0}}=\bra{(\bar{d}s)}
\end{array}\right]&=&
\left[\begin{array}{l}
\phantom{-}\bra{1,+1}\\
\phantom{-}\bra{1,\phantom{-}0}\\
-\bra{1,-1}
\end{array}\right]\,.
\end{eqnarray}
From this we derive the decomposition of the neutral states
\begin{align}
&\bra{\pi^0} &=\quad &-\frac{1}{2}\bra{1,0}+\frac{\sqrt{3}}{2}\bra{0,0}_8\,,\hfill\nonumber\\
&\bra{\eta}  &=\quad &\sqrt{\frac{2}{3}}\bra{1,0}+\frac{\sqrt{2}}{3}\bra{0,0}_8-\frac{1}{3}\bra{0,0}_1\,,\hfill\nonumber\\
&\bra{\eta' }&=\quad &\frac{1}{2\sqrt{3}}\bra{1,0}+\frac{1}{6}\bra{0,0}_8+\frac{2\sqrt{2}}{3}\bra{0,0}_1\,,\hfill\nonumber
\end{align}
\begin{align}
&\bra{\rho^0}&=\quad &-\frac{1}{2}\bra{1,0}+\frac{\sqrt{3}}{2}\bra{0,0}_8\,,\hfill\nonumber\\
\hspace{3cm}&\bra{\omega}&=\quad &-\frac{1}{2}\bra{1,0}-\frac{\sqrt{3}}{6}\bra{0,0}_8+\sqrt{\frac{2}{3}}\bra{0,0}_1\,,\hspace{3cm}\nonumber\\
&\bra{\phi}  &=\quad &\frac{1}{\sqrt{2}}\bra{1,0}+\frac{1}{\sqrt{6}}\bra{0,0}_8+\frac{1}{\sqrt{3}}\bra{0,0}_1\,,\hfill
\end{align}
where the subscript $1,8$ on the two \us singlet states refers to the $SU(3)$ transformation properties of
the corresponding state.

It is interesting to note that one may infer some relations in the \us limit. These have been discussed 
in the literatue \cite{FleischerUSpin,GronauUSpin}, but are rederived here in a different way.
The key observation is that due to CKM unitarity all CP violation in the standard model is 
proportional to the Jarlskog invariant
\begin{equation} \label{Delta} 
{\rm Im} \Delta = {\rm Im} (V_{cb} V_{cd}^* V_{ub}^* V_{ud}) = 
                          - {\rm Im} (V_{cb} V_{cs}^* V_{ub}^* V_{us}) \,.
\end{equation}  
In particular, all CP violating rate differences $\Delta \Gamma = \Gamma (B \to f) - \Gamma (\overline{B} \to \overline{f})$
are proportional to ${\rm Im} \Delta$. Exchanging the roles of the $d$ and the $s$ quark will 
flip the sign of ${\rm Im} \Delta$ in $\Delta \Gamma$ as can bee seen in (\ref{Delta}). 

The relation (\ref{Delta}) may be combined with the group theory of \usd. 
We note that we have for the operators in the effective hamiltonian 
\begin{equation} \label{Uplus} 
\left[ U_\pm  \, , \, Q^{(1/2)}_{\mp 1/2}\right] = Q^{(1/2)}_{\pm 1/2}\,,   \qquad  \left[U_\pm  \, , \, P^{(1/2)}_{\mp 1/2} \right] = P^{(1/2)}_{\pm 1/2} \,,
\end{equation}
where $U_\pm$ is the operator which raises or lowers the 3-component of the \us by one unit, respectively.

For the case of charged $B$ mesons we have a \us singlet in the initial state, and hence the final states 
may only have $U = 1/2$. Using (\ref{Uplus}) we see that 
\begin{equation} 
\langle B^+ | Q^{(1/2)}_{-1/2} | f, 1/2 \rangle =  \langle B^+ | Q^{(1/2)}_{+1/2} | f, -1/2 \rangle \,,
\end{equation} 
and the analogous relation  for the matrix elements of $P^{(1/2)}_{M}$. In the effective hamiltonian 
these matrix elements appear with CKM factors in which the role of the $s$ and $d$ quarks are interchanged 
and hence we get in he \us limit
\begin{equation}
\Delta \Gamma (B^+ \to (f, U_3 = 1/2)) = - \Delta \Gamma (B^+ \to (f, U_3 = -1/2))
\end{equation}
for any state $f$. 

The neutral $B$ mesons form a \us doublet, and hence the possible final states can be either a 
singlet or a triplet. Using the above reasoning, we infer the relation
\begin{equation}
\Delta \Gamma (B_d \to (f, U=0, U_3=0)) = - \Delta \Gamma (B_s \to (f, U=0, U_3=0)) \,.
\end{equation}

Finally, the case of a triplet final state yields also similar relations for the $U_3 = \pm 1$ 
components. Using again (\ref{Uplus}) we get 
\begin{equation}
\Delta \Gamma (B_d \to (f, U=1, U_3=1)) = - \Delta \Gamma (B_s \to (f, U=1, U_3=-1))\,.
\end{equation}

These relations may serve as a test for the amount of \us breaking. In fact, rewriting 
the relation for the rate differences into the usual observables we get   
\begin{equation}\label{ObsRelation}
\frac{A_{\rm CP}(\#1)}{A_{\rm CP}(\#2)}=-\frac{\Gamma(\#2)}{\Gamma(\#1)}\,,
\end{equation} 
which we shall check once we will discuss applications. However, one has to keep in mind, 
that these relations also reduce the number of independent observables. 

Finally we remark that we can supplement the \us relations by isospin symmetry.  As pointed out 
above, we consider it safe to use isospin as an exact symmetry in this context and use this to constrain the  
\us breaking parameters.  
The fundamental doublets  are defined as 
\begin{equation}
\left[\begin{array}{l}
|u\rangle \\
|d\rangle \end{array}\right]
=
\left[\begin{array}{l}
|\frac{1}{2}\,+\frac{1}{2}\rangle\\
|\frac{1}{2}\,-\frac{1}{2}\rangle\\
\end{array}\right]_I\,,\quad
\left[\begin{array}{l}
|\bar{d}\rangle \\
|\bar{u}\rangle \end{array}\right]
=
\left[\!\begin{array}{l}
\phantom{-}|\frac{1}{2}\,+\frac{1}{2}\rangle\\
-|\frac{1}{2}\,-\frac{1}{2}\rangle\\
\end{array}\right]_I\,,\\
\end{equation}
which fixes our conventions.  

However, under isospin the effective hamiltonian decomposes in a more  complicated way. The transformation 
properties of the operators defined in (\ref{IrrOp}) are given in table~\ref{IsospinClassification}. 
Note that for the $Q_{\pm 1/2}^{(1/2)}$ operators the only topology contributing with $\Delta I=1,3/2$ 
respectively is the electroweak penguin.

\mytab[IsospinClassification]{|l|l|}{\hline
Operator & $(\Delta I,\Delta I_z)$\\\hline\hline
$\Delta C=1:$&\\\hline
$P_{+1/2}^{(1/2)}$ & $(1,-1)$\\
$P_{-1/2}^{(1/2)}$ & $(1/2,-1/2)$\\\hline\hline
$\Delta C=0, b\to d$ &\\\hline
$Q_{+1/2}^{(1/2)}$ & $(1/2,-1/2)\oplus(3/2,-1/2)$\\
$R_{+1/2}^{(1/2)}$ & $(1/2,-1/2)\oplus(3/2,-1/2)$\\\hline\hline
$\Delta C=0, b\to s$ &\\\hline
$Q_{-1/2}^{(1/2)}$ & $(0,0)\oplus(1,0)$\\
$R_{-1/2}^{(1/2)}$ & $(0,0)\oplus(1,0)$\\\hline
}{Classification of irreducible \us operators in terms of isospin.}

\section{Applications}
In this section we apply the above formalism to non-leptonic two-body $B$ decays. 
Clearly we shall not discuss all possible decays here, rather we focus on two sample 
applications to check how far we can get without any restrictive ad-hoc assumptions. 

As stated above, the charged $B$ mesons are \us singlets and hence - due to the simple \us 
structure of the effective hamiltonian - the final state has to be 
a doublet or - including \us breaking - a quadruplet. Considering two-body decays, this corresponds 
to having one final-state meson in a \us doublet, while the other has to be either \us 
singlet or triplet. 

In the case where one of the final state mesons is in a triplet there is another complication. Since \us 
breaking is not that small, the mass eigenstates are quite different from the \us eigenstates, i.e. there is 
not even an approximate mass eigenstate corresponding to an  $s\bar{s}-d\bar{d}$ \us state. 
As already described above (see (\ref{conv}) ff.),  there are three mass eigenstates contributing 
to $U_z=0$, which all have to be taken into account.

The two neutral $B$ mesons form a doublet under \usd. In the \us limit the contributing final states 
have to form either a singlet or a triplet, while we can have also admixtures of $U=2$ once we include 
\us breaking. When considering two-body decays there are in total three possibilities. The decays 
into two charged final states necessarily have either $U=0$ or $U=1$, since the charged mesons form 
\us doublets. The neutral mesons form either \us singlets or triplets, in which case the 
two-body final states can have $U=0,1$ and 2. Clearly a final state with $U=2$ can be reached only 
through \us breaking. 

In the following we discuss the general possibilities to constrain \us breaking. Evidently the data  on 
$B_s$ decays is needed which will be available in the near future from LHC data. However, based on 
the current data one may already discuss \us breaking in 
some modes, e.g. $B \to J/\psi \,  (K\mbox{ or }\pi)$ (charged and 
neutral) and in the decays $B^\pm \to D \, (K^\pm\mbox{ or }\pi^\pm)$.  

As will become apparent in these example analyses, the precision of the data does not yet suffice to draw strong conclusions. For some modes,
this may change with LHCb, however, the modes including neutral light mesons in the final states will only be accessible with a sufficient precision at
a super flavour factory. In general, the parameters appearing in our expressions can be constrained meaningfully in the relevant range with the expected precision from future experiments. We have 
checked this by performing some sample fits within a simple ``future scenario''. The details,  however,  depend strongly on the mode, and in some cases
one still has to resolve discrete ambiguities.

\subsection{$\mathbf{U}$-Spin Breaking in \boldmath $B \to M_0 M_0$ and in $B \to $ \unboldmath CP-Eigenstates} 
When considering decays of neutral $B$ mesons into two neutral mesons, one has to deal with admixtures of 
\us multiplets. Using  the decomposition   
\begin{eqnarray}
\mathcal{H}_{eff}^{b\to d}\bra{\bar{B}_d} &=& -\frac{1}{\sqrt{2}}\bra{1,0}_{d,0}
             -\frac{1}{\sqrt{3}}\bra{1,0}_{d,\epsilon(3/2)}+\frac{1}{\sqrt{6}}\bra{1,0}_{d,\epsilon(1/2)}
             \nonumber \\
\label{HonB01} & & -\frac{1}{\sqrt{2}}\bra{0,0}_{d,0}
                               +\frac{1}{\sqrt{6}}\bra{0,0}_{d,\epsilon}-\frac{1}{\sqrt{3}}\bra{2,0}_{d,\epsilon}\,,\\
\mathcal{H}_{eff}^{b\to s}\bra{\bar{B}_s} &=& +\frac{1}{\sqrt{2}}\bra{1,0}_{s,0}
             -\frac{1}{\sqrt{3}}\bra{1,0}_{s,\epsilon(3/2)}+\frac{1}{\sqrt{6}}\bra{1,0}_{s,\epsilon(1/2)}
             \nonumber\\
& & -\frac{1}{\sqrt{2}}\bra{0,0}_{s,0}-\frac{1}{\sqrt{6}}\bra{0,0}_{s,\epsilon}+\frac{1}{\sqrt{3}}\bra{2,0}_{s,\epsilon}
\,,
\end{eqnarray}
\begin{eqnarray}
\mathcal{H}_{eff}^{b\to d}\bra{\bar{B}_s} &=& +\bra{1,+1}_{d,0}
               -\frac{1}{\sqrt{6}}\bra{1,+1}_{d,\epsilon(3/2)}-\frac{1}{\sqrt{3}}\bra{1,+1}_{d,\epsilon(1/2)}
               \nonumber\\
& & +\frac{1}{\sqrt{2}}\bra{2,+1}_{d,\epsilon}\,,\\
\mathcal{H}_{eff}^{b\to s}\bra{\bar{B}_d} &=& -\bra{1,-1}_{s,0}
                 -\frac{1}{\sqrt{6}}\bra{1,-1}_{s,\epsilon(3/2)}-\frac{1}{\sqrt{3}}\bra{1,-1}_{s,\epsilon(1/2)}
                 \nonumber\\
\label{HonB02}& & -\frac{1}{\sqrt{2}}\bra{2,-1}_{s,\epsilon}\,,
\end{eqnarray}
one may express all the amplitudes in terms of \us amplitudes. 
Doing this in full generality leads to a large number of independent \us amplitudes for $U_z=0$ final 
states already in the symmetry limit, and does in general not allow for a determination of all breaking 
amplitudes. One theoretical exception is given by decays $B^0\to P^0P^0$: when both final state particles
belong to the same multiplet, Bose symmetry forbids antisymmetric final states, leading to a reduction of
possible amplitudes\footnote{This fact has been overlooked in \cite{SoniSuprun}, the corrections to the 
corresponding decompositions are straight forward.}. However, this possibility remains a theoretical one,
because in order to perform this fit, all decays of this class would have to be measured time-dependently,
which seems not possible in the near future.

Choosing the subset of decays formed by $b\to s$ transitions of $B_d$-mesons, combined with $b\to d$ 
transitions of $B_s$-mesons \cite{SoniSuprun} results in 19 parameters facing up to 18 observables, 
therefore in this case one additional assumption is needed.

In any case, the current situation concerning the data is insufficient to perform such an analysis, since 
the $B_s$ system has not been fully explored yet. Clearly with the advent of LHC this situation will change
once LHCb measures the decay rates and the CP asymmetries of the corresponding $B_s$ transitions.  

Decays into CP eigenstates (or, more generally, states which are not flavour-specific) play an exceptional role, 
because of the additional information coming from time-dependent measurements. Each of these decays forms a subset 
with its \us partner formed by exchanging all down and strange quarks in the process, because they have effectively 
only one amplitude. These subsets can be discussed separately from the rest of the corresponding class, which allows 
for fits with a small number of parameters, even when other decays of that class have not been measured yet. This 
feature has been extensively exploited in the \us limit, or 
including factorizable \us breaking only (see e.g. \cite{FleischerUSpin,GronauUSpin}). In that case, the two decays 
in question have five independent observables 
(because of relation (\ref{ObsRelation})), but only three parameters, so a fit for up to two weak phases is possible. 
However, these determinations suffer from the systematic uncertainty related to \us breaking.

Including the breaking corrections to first order for these subsets, one observes that the breaking amplitudes form 
only one effective breaking amplitude as well. However, again this does not suffice for an analysis of the breaking
which is completely free from additional assumptions: the number of parameters increases by four, while only one 
additional independent observable becomes available. In these cases, for example the following two strategies may be used:
\begin{itemize}
\item If one amplitude is clearly dominating ($|A_1/A_2|\sim\delta$), one may consider the \us breaking for the leading amplitude only, neglecting only terms of order ($\mathcal{O}(\epsilon^2),\mathcal{O}(\epsilon \delta)$). This is for example the case in $B\to D_{d,s}^+D_{d,s}^-$ decays, which are dominated by their colour allowed tree contribution.
\item If one of the two parts in the leading amplitude is dominated by a colour allowed tree contribution, one may use the factorization assumption for that part only, as opposed to using it for the whole amplitude, and fit for the breaking amplitude in the other part. 
\end{itemize}
In both cases, the number of free parameters increases only by two, so in principle a fit becomes possible; in addition, as one additional observable is available, one may determine that way one of the weak phases with correspondingly smaller systematic uncertainty. If for one class of decays the whole set is measured, these strategies may be used with the whole set, so the decays into flavour-specific modes can be included.

Finally, let us comment on the phenomenologically important decays $B_{d,s}\to(\pi/K)^+(\pi/K)^-$, of which a subsection are CP eigenstates. A priori, both strategies are not applicable in this case. However, there exists a small amplitude combination, corresponding to the fact that the amplitudes for $B_d\to K^+K^-$ and $B_s\to\pi^+\pi^-$ are ``dynamically suppressed'', i.e. they proceed only via  annihilation. This fact might be used by setting to zero the corrections to this amplitude combination with $\Delta I=1/2$, thereby reducing the number of parameters to 15, while there are in principle 16 measurements available from these decays. However, a fit still requires the time-dependent measurements of all these decays. At the moment, there is no sign of large \us breaking in these decays \cite{JungPhD}.

\subsection{The decays $\boldsymbol{B \to J/\psi\, (K  \mbox{ or }\pi)}$}
The decays $B \to J/\psi \, (K \, \mbox{or} \, \pi) $ are under the simplest cases of $\Delta C = 0$
from the group theoretical point of view, because of $J/\psi$ and $B^-$ being \us singlets.
Our analysis is based on the data shown in  tables \ref{Bcharged} and \ref{Bneutral}.
\mytab[Bcharged]{|c|c|c|}{\hline
Decay                           & ${\rm BR}/10^{-4}$                  & $A_{\rm CP}$                      \\ \hline
$B^-\to J/\psi K^-$             & $10.07\pm0.35$               & $0.017\pm0.016(*)$ \\
$B^-\to J/\psi \pi^-$           & $0.49\pm0.06(*)$   & $0.09\pm0.08$     \\
\hline
}
{Measurements for the decays $B^- \to J/\psi (K\mbox{ or }\pi)$, data taken from the
PDG \cite{PDG}. $(*)$: Error enhanced by the PDG because of inconsistent measurements.}
As a first step, we check for \us violation by testing the \us relation (\ref{ObsRelation}). 
Inserting the data from table \ref{Bcharged} and neglecting a tiny phase space difference, we get 
\begin{equation}\label{ObsRelPsiKpi}
( A_{\rm CP} \times {\rm BR})_{B^-\to J/\psi K^-}+( A_{\rm CP} \times {\rm BR})_{B^-\to J/\psi \pi^-}=0.22\pm0.17\,,
\end{equation}
adding errors simply in quadrature. This result is not significant and a real test may only be 
performed, if at least one of the  asymmetries is measured significantly different from zero. 

In many applications naive factorization has been applied, which allows to include at least the 
factorizable part of \us breaking. In this picture one expects the ratio of branching ratios 
to be given only in terms of CKM factors  and the ratio of form factors. One gets the theoretical 
prediction 
\begin{equation}
\frac{{\rm BR}(B^-\to J/\psi K^-)}{{\rm BR}(B^-\to J/\psi \pi^-)} \sim %
\left(\frac{F^{B\to K}(M_{J/\psi}^2)}{F^{B\to\pi}(M_{J/\psi}^2)}\right)^2%
\left|\frac{V_{cb}^* V_{cs}}{V_{cb}^* V_{cd}}\right|^2 =  33.9\pm6.1\,, 
\end{equation}
where the form factor ratio is taken from QCD sum rule calculations \cite{QCDSR} and scaled to
$q^2=m_{J/\psi}^2$ with aid of a simple BK ansatz \cite{BKParametrization}.
This has to be contrasted with the experimental number   
\begin{equation}
\frac{{\rm BR}(B^-\to J/\psi K^-)}{{\rm BR}(B^-\to J/\psi \pi^-)} = \left\{ \begin{array}{ll}
19.2\pm1.5 \quad \mbox{(measurement of the ratio)\,,} \\  21.4\pm1.9 \quad \mbox{(combined single measurements).} 
\end{array} \right.   
\end{equation}
The sizable discrepancy indicates the well 
known fact that  theses decays have large non-factorizable contributions. 

\begin{figure}
\begin{center}
\begin{minipage}{6cm}
\centering{$\delta_0\in[-\pi/2,\pi/2]$}\\
\includegraphics[scale=0.3]{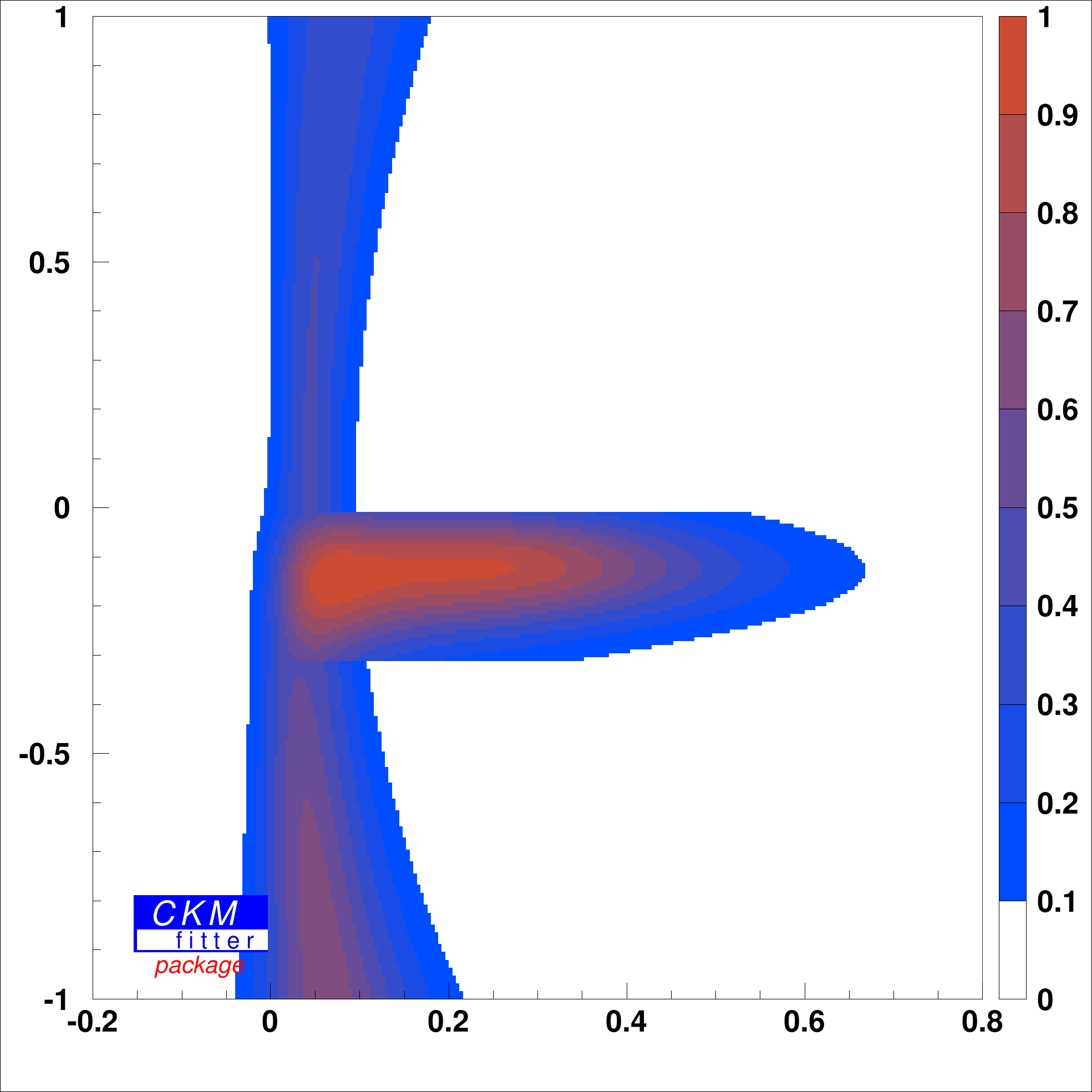}
\end{minipage}\qquad\qquad
\begin{minipage}{6cm}
\centering{$\delta_0\in[\pi/2,3\pi/2]$}\\
\includegraphics[scale=0.3]{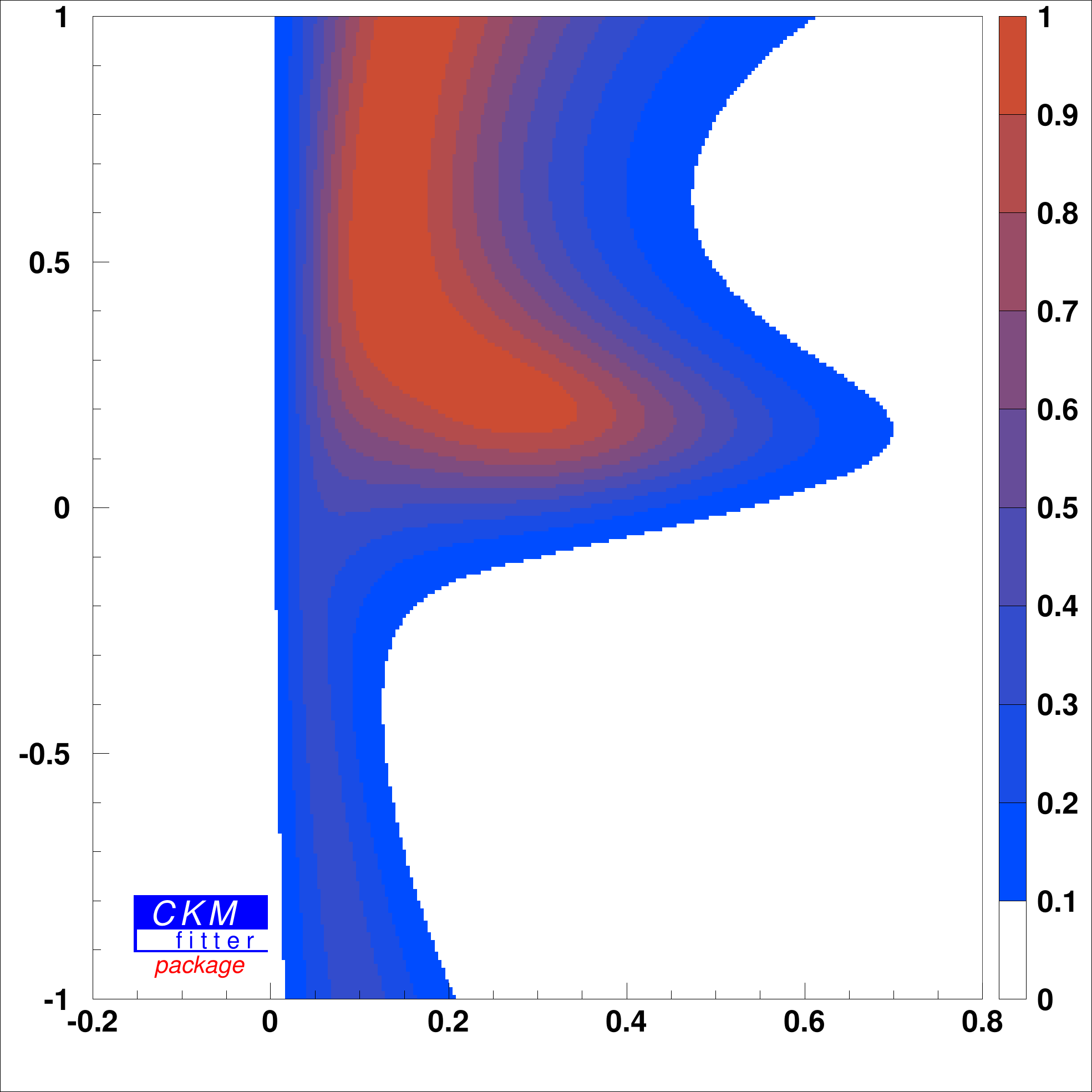}
\end{minipage}
\end{center}
\caption{The fit results for the \us breaking parameter $x_\epsilon$ in $B \to J/\psi K$ and $B \to J/\psi \pi$ in the complex plane.}
\label{fig1}
\end{figure}

On the other hand, the data in table \ref{Bcharged} are not sufficient to allow a fit to the general 
group theoretical expressions. Hence some additional assumptions are  necessary. We shall assume the 
following: 
\begin{itemize}
\item The amplitude proportional to $\lambda_{ud/s} = V_{ub} V_{ud/s}^* $ 
is expected to be small compared to the one proportional to $\lambda_{cd/s} = V_{cb} V_{cd/s}^* $,
because its tree operator contribution has only penguin matrix elements. Hence we will not take into account 
\us breaking for this amplitude.
\item We shall also make use of isospin symmetry. This means that we have to take into account 
also the decays of the neutral $B$ modes, since they are the isospin partners of the charged 
$B$ mesons. When making use of isospin, the matrix elements identified in the \us analysis are splitted into their two
isospin components as shown in table \ref{IsospinClassification}. Here we neglect the contribution with $\Delta I=1,3/2$ 
proportional to $\lambda_{cs/d}$, which receive contributions from penguin matrix elements of electroweak penguin operators only; 
hence  we assume the corresponding penguin contributions to be a pure $\Delta I=0,1/2$ amplitude for both the $b\to s$ and 
$b \to d$ transition.
\end{itemize}
\mytab[Bneutral]{|c|c|c|c|}{\hline
Decay                           & ${\rm BR}/10^{-4}$                  & $A_{\rm CP}$                      & $S_{\rm CP}$\\\hline
$\bar{B}^0\to J/\psi \bar{K}^0$ & $8.71\pm0.32$      & $-0.002\pm0.020(*)$           & $0.657\pm0.025$\\
$\bar{B}^0\to J/\psi \pi^0$     & $0.205\pm0.024$    & $0.10\pm0.13$      & $-0.93\pm0.29(**)$\\\hline%
}
{Measurements for the decays $\bar{B}\to J/\psi (K\mbox{ or }\pi)$. Time-dependent measurements are taken from the HFAG 
\cite{HFAG}, other data from the PDG \cite{PDG}. $(*)$: Error enhanced by the PDG due to inconsistent measurements. 
$(**)$: Error enhanced according to the PDG prescription for the same reason.}
For the neutral $B$ mesons we include the data shown in table~\ref{Bneutral}. Using the above 
assumptions, we are lead to the following parametrization:
\begin{eqnarray} \label{Bparm} 
<B^-|\mathcal{H}_{eff}|J/\psi K^->             &=& N_{J/\psi K}\left(1+x_\epsilon+\epsilon\, e^{-i\gamma}\,r_0\,e^{i\phi_0}\right)\,,\nonumber\\
<B^-|\mathcal{H}_{eff}|J/\psi \pi^->           &=& \frac{\lambda}{1-\lambda^2/2} N_{J/\psi K}\left(-1+x_\epsilon+e^{-i\gamma}\,r_0\,e^{i\phi_0}\right)\,,\nonumber \\
<\bar{B}^0|\mathcal{H}_{eff}|J/\psi \bar{K}^0> &=& N_{J/\psi K}\left[1+x_\epsilon+\epsilon\, e^{-i\gamma}\left(r_0\,e^{i\phi_0}-2r_1^K\,e^{i\phi_1^K}\right)\right]\,,\nonumber\\ 
<\bar{B}^0|\mathcal{H}_{eff}|J/\psi \pi^0>     &=& \frac{\lambda}{1-\lambda^2/2}N_{J/\psi K}\left[-1+x_\epsilon+e^{-i\gamma}\left(r_0\,e^{i\phi_0}-2r_{3/2}^\pi\,e^{i\phi_{3/2}^\pi}\right)\right]\,,
\end{eqnarray}
where the normalization factor $N_{J/\psi K}$ is chosen such that $N_{J/\psi K}^2={\rm BR}(B^-\to J/\psi K^-)$ 
in absence of \us breaking and penguin effects, which implies 
$\epsilon=|V_{ub}V_{us}^*|/|V_{cb}V_{cs}^*|$. As a consequence, the corresponding ratios of lifetimes 
and phase space factors have to be taken into account when computing the branching ratios from 
(\ref{Bparm}) for the other decays. Furthermore, the ratios $r_0$, $r_1^K$ and $r_{3/2}^\pi$ are 
the penguin and $u$ quark tree contributions (normalized to  $N_{J/\psi K}$) respectively, which contain a factor 
$R_u = |V_{ub} V_{ud}^* / V_{cb} V_{cd}^* |$. Finally, the complex parameter $x_\epsilon$ represents the 
\us breaking part in the leading contribution, again  normalized to  $N_{J/\psi K}$. 
As inputs from the CKM fit we use, in addition to the ones described above, those from table~\ref{tableCKMdata}.
\mytab[tableCKMdata]{|c|c|}{\hline
Parameter            & Global fit value\\\hline
$\lambda$            & $0.2252\pm0.0008$\\
$\gamma$             & $\left(66.8^{+5.4}_{-3.8}\right)^\circ$\\
$\beta_{{\rm w/o} J/\psi}$ & $0.48^{+0.02}_{-0.04}$\\\hline}
{CKM parameters taken from \cite{CKMfitter}, results as of summer 08. The lower uncertainties of $\gamma$ and $\beta_{\mbox{w/o} J/\psi}$,
which refers to the fit to $\beta$ excluding the measurement of $\sin(2\beta)$ from $B\to J/\psi K_S$, have been slightly enhanced to reflect the 
non-gaussian behaviour of the distribution in a conservative way.}
The results of our fit are given in table \ref{resultsJPsipiK}, the results for the \us breaking parameter are additionally 
shown shown in fig.~\ref{fig1}.  The fit shows three distinct solutions, two of which 
have $\phi_0\sim 0$, while the third one has $\phi_0\sim\pi$. As the solutions interfere in the fit and make 
it unstable, we perform two separate fits with the restrictions $\phi_0\in[-\pi/2,\pi/2]$ and $\phi_0\in [\pi/2,3\pi/2]$, 
covering the whole parameter space.

\refstepcounter{tables}%
\begin{table}[t!hb]%
\begin{center}%
\begin{tabular}{|c|c|c|c|}%
\hline
\multicolumn{4}{|c|}{$\phi_0\in[-\pi/2,\pi/2]$ ($\chi^2=0.51$)}\\\hline
Parameter         & best fit value    & $1\sigma$ range                                                        & $2\sigma$ range \\\hline
$\mbox{Re}(x_\epsilon)$ & \phantom{-}0.08   & $[\phantom{-}0.02,0.41]$                                    & $[-0.03,\phantom{1}0.63]$\\
$\mbox{Im}(x_\epsilon)$ & -0.14             & $[-0.28,-0.04]\; \vee \leq-0.6$                                          & unconstrained\\
$N_{J/\psi K}^2$  & \phantom{-}8.39   & $[\phantom{-}4.60,9.38]$                                    & $[\phantom{-}3.65,10.17]$\\
$r_0$             & \phantom{-}0.88   & $[\phantom{-}0.07,\phantom{-}0.26] \vee [\phantom{-}0.56,1.47]$        & $[\phantom{-}0.00,\phantom{1}1.72]$\\
$\phi_0$          &  \phantom{-}0.09  & $[-0.22,0.61]$                                              &unconstrained\\
$r_1^K$           & \phantom{-}1.60   & $[\phantom{-}1.18,2.37]$                                    & $[\phantom{-}0.66,\phantom{1}2.85]$\\
$\phi_1^K$        & -0.07             & $[-0.75,-0.50] \vee [-0.17,0.04]$                                      & $[-0.90,\phantom{1}0.88]$\\
$r_{3/2}^\pi$     & \phantom{-}0.49   & $[\phantom{-}0.00,\phantom{-}0.09]\vee[\phantom{-}0.24,1.18]$          & $[\phantom{-}0.00,\phantom{1}1.46]$\\
$\phi_{3/2}^\pi$  & \phantom{-}(0.16) & unconstrained                                                          & unconstrained\\\hline\hline

\multicolumn{4}{|c|}{$\phi_0\in[\pi/2,3\pi/2]$ ($\chi^2=0.01$)}\\\hline
Parameter         & best fit value & $1\sigma$ range               & $2\sigma$ range \\\hline
$\mbox{Re}(x_\epsilon)$ & 0.13           & $[0.06,0.45]$      & $[\phantom{-}0.01,0.66]$\\
$\mbox{Im}(x_\epsilon)$ & (0.59)         & $\geq0.06$      & unconstrained\\
$N_{J/\psi K}^2$  & 6.27           & $[3.76,8.60]$ & $[\phantom{-}2.96,9.93]$\\
$r_0$             & 0.29           & $[0.09,1.03]$      &$[\phantom{-}0.00,1.38]$\\
$\phi_0$          & 2.78           & $[2.28,3.25]$      & unconstrained\\
$r_1^K$           & 1.40           & $[0.78,2.14]$      &$[\phantom{-}0.31,2.58]$\\
$\phi_1^K$        & 0.57           & $[0.05,0.91]$      &$[-0.87,1.11]$\\
$r_{3/2}^\pi$     & 0.06           & $[0.00,0.18]$      &$[\phantom{-}0.00,0.31]$\\
$\phi_{3/2}^\pi$  & (2.55)         & unconstrained      &unconstrained\\\hline
\end{tabular}%
\end{center}%
\caption{\label{resultsJPsipiK}\small Results for the fit to $J/\psi (K\mbox{ or }\pi)$ data, as explained in the text. The values in brackets indicate that due to a broad allowed range the central value is not significant.}
\end{table}

The sizable difference between the branching ratios of the charged and the neutral $B \to J/\psi K$ modes 
is somewhat surprising. The isospin analysis shows that it is driven 
by the $\Delta I = 1$ contribution of the effective hamiltonian, which is doubly CKM suppressed. Hence    
the ratio between these branching ratios should be given by the ratio of lifetimes which is close to unity, 
modified only by a doubly Cabbibo suppressed tree contribution and electroweak penguins. 
In our fit, this results in $r_1^K\sim1 (\geq0.78@1\sigma)$, which is quite large, but on the other hand not
conclusive at the moment.
Furthermore, the non-vanishing central values for the CP asymmetries imply a non-vanishing value for $r_0$, 
$r_0\ge0.09 @ 1\sigma$, in combination with a non-trivial phase. 
However, as is obvious from the significance
of the data, the allowed range at two standard deviations includes zero.
Concerning \us breaking, the fit prefers a 
non-vanishing imaginary part of the \us breaking parameter $x_\epsilon$, while it is not bounded from above. 
The first observation is due to (\ref{ObsRelPsiKpi}) 
showing a deviation from zero, and especially preferring equal signs for the CP asymmetries, while the branching 
ratios, as seen above, are compatible with no breaking at all. This is again a hint to non-factorizable \us breaking. 
The reason for the second observation lies in the fact, that no observable depends in leading order on 
$\mbox{Im}(x_\epsilon)$ when assuming a power-counting $x_\epsilon,r_i\sim \lambda$. 
The other parameters lie within relatively large ranges, whithin or including the expected order of magnitude.

\subsubsection{\boldmath $B^\pm\to D \, ( K^\pm \, \mbox{or} \, \pi^\pm)$ \unboldmath decays}
As an example for $\Delta C = \pm1$ transtions we consider the decays 
$B^\pm\to D \, ( K^\pm \, \mbox{or} \, \pi^\pm)$.
As mentioned above, these transitions are governed 
by a single CKM factor, since there are four different quark flavours in the final state. 
In particular, this leads to vanishing direct CP asymmetries in the corresponding decays, so 
the number of parameters as well as the one of observables is less by a factor of two. 

However, it has been proposed a few years ago \cite{GronauWyler,GronauLondon} to discuss observables from decays, 
where the (neutral) D meson in the final state is reconstructed in a decay mode which is a CP eigenstate.  
This leads to interference between  $B\to D$- and $B\to\bar{D}$-modes, where $B^-\to D K^-$ is the 
``golden mode'' to extract $\gamma$ with negligible theoretical error. The analysis can be transferred 
to $B\to D\pi$ one to one, however, in this case the second amplitude, $B^-\to \bar{D}^0\pi^-$ is not only 
colour-, but in addition doubly Cabibbo-suppressed, 
which leads to very small interference effects.

Turning to \usd, the analysis is analogous to the $B \to J/\psi (K \, \mbox{or} \, \pi)$ modes, 
however, with only a single CKM factor in each amplitude. The parametrization in this case reads
\begin{eqnarray} \label{BDKparm}
\langle B^-|\mathcal{H}_{eff}|D^0K^-\rangle &=& 
\lambda \tilde{A}_1\left(1+y_{1,\epsilon} e^{i\theta_1}\right)\,,\nonumber\\
\langle B^-|\mathcal{H}_{eff}|D^0\pi^-\rangle &=& 
(1-\lambda^2/2)\tilde{A}_1\left(1-y_{1,\epsilon} e^{i\theta_1}\right)\,,\nonumber\\
\langle B^-|\mathcal{H}_{eff}|\bar{D}^0K^-\rangle &=& 
\lambda R_ue^{-i\gamma} \tilde{A}_2 e^{i\theta_A}\left(1+ y_{2,\epsilon} e^{i\theta_2}\right) \,, 
\nonumber\\
\langle B^-|\mathcal{H}_{eff}|\bar{D}^0\pi^-\rangle &=& 
-\lambda^2 R_ue^{-i\gamma} \tilde{A}_2 e^{i\theta_A}\left(1-y_{2,\epsilon} e^{i\theta_2}\right)\,,
\end{eqnarray}
a common factor $A\lambda^2$ is absorbed into the definition 
of $\tilde{A}_{1,2}$. In the fit, we include the (to order $1\%$)  common phase-space factor $\Phi$ 
and the lifetime of the $B$-meson by the definition
\begin{equation}
A_{1,2}=\sqrt{\Phi(m_B,m_{\pi})\tau_{B^-}}\tilde{A}_{1,2}\,.
\end{equation}
Note that we choose $A_1$ to be real, while for $A_2$ one has to keep a phase because of the interference 
effects described below. Furthermore, the \us breaking quantities $y_{1,\epsilon}$ and 
$y_{2,\epsilon}$ 
are real and positive, since their phases are taken into account explicitely.   

Defining now the CP eigenstates\footnote{In the follwing we neglect any mixing in the $D$ system.} 
\begin{equation}
|D^0_\pm\rangle=\frac{1}{\sqrt{2}}\left(|D_0\rangle\pm|\bar{D}_0\rangle\right)\,,
\end{equation}
one has the additional observables
\begin{eqnarray}
\bar{\Gamma}(B^-\to D_\pm^0 K^-/\pi^-)&=&\frac{1}{2}\left(\Gamma(B^-\to D_\pm^0 K^-/\pi^-)+\Gamma(B^+\to D_\pm^0 K^+/\pi^+)\right)\,, \\
A_{\rm CP}(B^-\to D_\pm^0 K^-/\pi^-)&=&\frac{\Gamma(B^-\to D_\pm^0 K^-/\pi^-)-\Gamma(B^+\to D_\pm^0 K^+/\pi^+)}{\Gamma(B^-\to D_\pm^0 K^-/\pi^-)+\Gamma(B^+\to D_\pm^0 K^+/\pi^+)}\,,
\end{eqnarray}
with the four relations 
\begin{eqnarray}
\label{ACPrel}
&& \bar{\Gamma}(B^-\to D_+^0 K^-/\pi^-)\mathcal{A}_{\rm CP}(B^-\to D_+^0 K^-/\pi^-)= \nonumber \\ 
&& \qquad \qquad -\bar{\Gamma}(B^-\to D_-^0 K^-/\pi^-)\mathcal{A}_{\rm CP}(B^-\to D_-^0 K^-/\pi^-)\,, \\[1.2ex]
\label{Gammarel}
&& \bar{\Gamma}(B^-\to D_+^0 K^-/\pi^-)+\bar{\Gamma}(B^-\to D_-^0 K^-/\pi^-)=\nonumber  \\
&& \qquad \qquad \bar{\Gamma}(B^-\to D^0 K^-/\pi^-)+\bar{\Gamma}(B^-\to \bar{D}^0 K^-/\pi^-)\,,
\end{eqnarray}
where the first relations require in particular opposite signs for the CP-asymmetries. Furthermore, in the \us
limit, relation~(\ref{ObsRelation}) implies
\begin{equation}\label{uslimitDpiK}
A_{\rm CP}(B^-\to D_\pm^0 K^-) {\rm BR}(B^-\to D_\pm^0 K^-)=-A_{\rm CP}(B^-\to D_\pm^0 \pi^-) {\rm BR}(B^-\to D_\pm^0 \pi^-)\,.
\end{equation}
Including \us breaking, this leaves eight independent observables in total.

The eight observables face 7 parameters appearing in (\ref{BDKparm}), if the weak angle $\gamma$ is 
treated as an input, otherwise we have to deal with 8 parameters. 
However, one has to take into account parametric invariances: one 
observes one discrete invariance\footnote{$\gamma$ has been restricted to lie in $[0,\pi]$, which excludes 
additional solutions.},
\begin{equation}
\gamma\to\pi-\gamma\,,\quad \theta_A\to\pi-\theta_A\,,\quad \theta_{1,2}\to-\theta_{1,2}\,,
\end{equation}
which leaves all observables invariant because this transformation effectively replaces every phase by 
its negative value. In the future, as long as $\gamma$ does not lie 
near $90^\circ$ (which is not the case, according to present data), this ambiguity is trivially resolved by 
the observation of other $\gamma-$dependent processes.

In addition there is  one continuous invariance: One has the freedom to redefine the parametrization
(\ref{BDKparm})  in such a way, that
\begin{eqnarray}
A_{1,2}(1+ y_{1,2}e^{i\theta_{1,2}})&\to&
A_{1,2}^{'}(1+ y_{1,2,\epsilon}^{'}e^{i\theta_{1,2}^{'}})=%
e^{i\theta_{\xi}^1}A_{1,2}(1+ y_{1,2}e^{i\theta_{1,2}})\,,\quad\mbox{and}\\
A_{1,2}(1- y_{1,2}e^{i\theta_{1,2}})&\to& 
A_{1,2}^{'}(1- y_{1,2,\epsilon}^{'}e^{i\theta_{1,2}^{'}})=%
e^{i\theta_\xi^2}A_{1,2}(1- y_{1,2}e^{i\theta_{1,2}})\,,
\end{eqnarray}
which is always possible in a restricted range for $\theta_\xi^{i}$. The restriction is given by the 
possible values of the corresponding parameter combinations, when considering $y_{1,2}\in[0,0.6]$ in the fit.

The experimental results for these decays are given in table \ref{tableBDKpidata}. The two colour 
suppressed decays have not been measured so far, the two CP asymmetries $B\to D^0 K^-/\pi^-$ 
do not enter the fit, because they are zero by construction, but are given mainly for completeness. 
Note that they are consistent with zero at the 1- and 2-sigma level respectively. 

We observe that the data are only roughly consistent with relations~(\ref{ACPrel}), within two standard
deviations. In addition, using relations~(\ref{Gammarel}), one observes 
that while the data for $B\to D K$ seems reasonable, the data for $B\to D\pi$ prefer a vanishing 
colour-suppressed amplitude by giving a negative central values for it. While this is on one hand sensible, 
because this amplitude is expected to be small, it is at odds with the measured non-vanishing CP-asymmetries. 
Together, these observations lead to a bad $\chi^2_{min}$-value in a global fit to the experimental data, 
independent of the \us breaking parameters.

\mytab[tableBDKpidata]{|c|c|}{\hline
Observable                                                         & Value\\\hline
${\rm BR}(B^-\to D^0 \pi^-)$                                       & $(48.4\pm 1.5)\,10^{-4}$\\
$A_{\rm CP}(B^-\to D^0 \pi^-)$                                     & $-0.008\pm0.008$\\
$\frac{{\rm BR}(B^-\to D^0 K^-)}{{\rm BR}(B^-\to D^0 \pi^-)}(*)$   & $0.076\pm 0.006$\\
$A_{\rm CP}(B^-\to D^0 K^-)$                             & $0.07\pm0.04$\\
$\frac{2{\rm BR}(B^-\to D^0_+ K^-)}{{\rm BR}(B^-\to D^0 K^-)}$     & $1.10\pm0.09$\\
$A_{\rm CP}(B^-\to D^0_+ K^-)$                           & $0.24\pm0.07$\\
$\frac{2{\rm BR}(B^-\to D^0_- K^-)}{{\rm BR}(B^-\to D^0 K^-)}$     & $1.06\pm0.10$\\
$A_{\rm CP}(B^-\to D^0_- K^-)$                           & $-0.10\pm0.08$\\
$\frac{{\rm BR}(B^-\to D^0_+ K^-)}{{\rm BR}(B^-\to D^0_+ \pi^-)}$  & $0.086\pm 0.009$\\
$A_{\rm CP}(B^-\to D^0_+ \pi^-)$                         & $0.035\pm0.024$\\
$\frac{{\rm BR}(B^-\to D^0_- K^-)}{{\rm BR}(B^-\to D^0_- \pi^-)}$  & $0.097\pm 0.017$\\
$A_{\rm CP}(B^-\to D^0_- \pi^-)$                         & $0.017\pm0.026$\\\hline
}
{Experimental data for $B^-\to D K^-/\pi^-$ decays. Data for $B\to D_\pm K$ is taken from the HFAG \cite{HFAG}, the rest from the PDG \cite{PDG}. 
$(*)$: Error rescaled by the PDG, due to inconsistent measurements.}

Checking now in a next step for \us breaking by evaluating relations~(\ref{uslimitDpiK}), we find good agreement
in case of the data for $B\to D^0_- (K\mbox{ or }\pi)$, while the relation for the $D^0_+$ data shows significant
\us violation, because both CP asymmetries are significantly different from zero and have the same sign. Therefore
we expect non-vanishing \us breaking parameters in the corresponding fit.

It is interesting to note that for the colour allowed tree decays one may check again naive factorization. In this 
case the \us breaking is given by the ratio of the decay constants, i.e. 
\begin{equation}
\frac{\langle B^-|\mathcal{H}_{eff}|D^0K^-\rangle}{\langle B^-|\mathcal{H}_{eff}|D^0\pi^-\rangle} = 
\frac{\lambda}{ 1-\frac{\lambda^2}{2}} \left(\frac{1+y_{1,\epsilon} e^{i\theta_1}}{1-y_{1,\epsilon} e^{i\theta_1}}\right)
\simeq \frac{\lambda}{ 1-\frac{\lambda^2}{2}}\frac{f_K}{f_\pi}\sim0.28\, . 
\end{equation}
In this approach we obtain $\theta_1 = 0$ and $y_{1,\epsilon} \sim 0.1$ from the ratio of the 
decay constants. The comparison with experiment (see table~\ref{tableBDKpidata}),
\begin{equation}
\sqrt{\frac{{\rm BR}(B^-\to D^0K^-)}{{\rm BR}(B^-\to D^0\pi^-)}}=0.276\pm0.011\,,
\end{equation}
shows excellent agreement, indicating the well known fact that naive factorization works reasonably
well in colour-allowed tree decays.

One may use this observation to fix $\theta_1\equiv0$, thereby breaking the parametric invariances described
above. This results in $\chi^2/{\rm d.o.f.}=7.42/4$, but mostly independent of the assumption concerning \us 
breaking, corresponding to the above discussion. While the results are therefore to be handled with care, 
we note that the fit prefers large values \us breaking parameter $r_2$, and leads to the predictions 
${\rm BR}(B^-\to\bar{D}^0 K^-)\in[0.02,0.09]\times10^{-4}$ and 
${\rm BR}(B^-\to\bar{D}^0 \pi^-)\leq0.03\times10^{-4}$ at $1\sigma$.

\section{Conclusions} 
Since methods based on factorization do not seem to converge quickly to allow 
for a reliable prediction for non-leptonic decays, the method of flavour symmetries looks 
more promising. Clearly the latter will allow us to perform precision calculations only 
if we get a reasonable control over symmetry breaking. 

Using the full $SU(3)$ flavour symmetry becomes quite complicated once its complete breaking 
is taken into account. However, the isospin subgroup of full $SU(3)$ may be 
assumed  to be a reasonably good symmetry and hence only the breaking along the 
``orthogonal'' directions in $SU(3)$ space has to be considered. 

We have studied the \us subgroup of $SU(3)$, which has the advantage that the 
charge operator commutes with the symmetry generators and hence also the weak 
hamiltonian for $B$ decays has a simple structure under this symmetry. The breaking 
term is due to the mass difference between the down and the strange quark and 
has a simple structure inferred from QCD. 

Based on this we have discussed how \us breaking can be incorporated on a purely 
group theoretical basis. We have shown a few applications, in which the \us breaking 
tuns out roughly of the order implied by the difference in the decay constants 
$f_\pi$ and $f_K$.

However, the full strength of this strategy can be exploited only in the future. Since the $B_d$ and the $B_s$ form 
a \us doublet,  the approach requires information on decay modes which will be gathered in 
the near future at  the LHC. With a sufficient amount of data there will be a chance to obtain 
control over flavour $SU(3)$ breaking and hence a possible road to precise predictions 
for non-leptonic decays may be opened.

\subsection*{Acknowledgements}
This work was supported by the german research foundation DFG under \\
contracts no. MA1187/10-1 and KH205/1-1 and by the EU
MRTN-CT-2006-035482 (FLAVIAnet), by MICINN (Spain) under grant
FPA2007-60323, and by the Spanish Consolider-Ingenio 2010
Programme CPAN (CSD2007-00042). 

\bibliography{paper_final.bib}
\end{document}